\documentclass[conference]{IEEEtran}
\usepackage{cite}
\usepackage{amsmath,amssymb,amsfonts}
\usepackage{graphicx}
\usepackage{textcomp}
\usepackage{xcolor}
\usepackage{url}
\usepackage{hyperref}

\title{DHMS: A Digital Hostel Management System Integrating Campus ChatBot, Predictive Intelligence, and Real-Time Automation}

\author{
\IEEEauthorblockN{Riddhi Heda, Sidhant Singh, Umair Yasir, Tanmay Jaiswal, Anil Mokhade}
\IEEEauthorblockA{\textit{Department of Computer Science and Engineering} \\
\textit{Visvesvaraya National Institute of Technology, Nagpur, India} \\
\{bt21cse001, bt21cse014, bt21cse068, bt21cse107\}@students.vnit.ac.in, asmokhade@cse.vnit.ac.in}
}

\begin{document}
\maketitle

\begin{abstract}
\begin{abstract}
Traditional manual hostel management practices continue to prevail in academic institutions, typically being plagued by inefficiencies, procrastination, and piecemeal communication. Such traditional systems don't live up to the standards of digitally native students and impose excessive operational loads on hostel staff. This paper presents DHMS (Digital Hostel Management System), an integrated, modular platform for digitizing and streamlining all fundamental functions of hostel management. In simulation runs, DHMS achieved 92\% student satisfaction in room allocation and an average chatbot response time under 1 second. DHMS utilizes contemporary web technologies, artificial intelligence, and cloud infrastructure for automating room allotment, grievance redressal, gate pass logistics management, and direct communication using a natural language chatbot. The system also has predictive analytics to plan maintenance proactively and sentiment analysis for processing feedback\textsuperscript{\cite{bert}}, thus improving the responsiveness of the institution. To move from prototype to full production, DHMS still requires end-to-end integration testing across multiple hostel blocks, user acceptance trials, load-testing for scale, and ERP-system integration before campus-wide deployment. DHMS architecture, approach, and deployment factors are discussed here in light of enhancing user experience, administration effectiveness, and decision-making astuteness.
\end{abstract}

\end{abstract}

\begin{IEEEkeywords}
Hostel Automation, ChatBot, Predictive Maintenance, Sentiment Analysis, AWS Lex, WebSocket, MERN Stack, Role-Based Access Control, NLP
\end{IEEEkeywords}

\section{Introduction}
Hostel management on university campuses has a significant influence on students' emotions and daily functioning. Yet, in most Indian institutions, hostel systems remain antiquated—paper registers for grievances, handwritten room lists, and leave applications requiring multiple physical signatures. Existing campus ERP platforms often provide broad functionality but tend to be monolithic, hard to customize, and lack deeper AI-driven features, while standalone hostel systems may be modular but rarely integrate real-time automation, predictive analytics, and conversational interfaces in one solution. These traditional practices slow down the process, cause confusion, and make both students and staff unhappy because of the lack of transparency and real-time information.

Whereas technology has been extensively used for areas such as study and e-learning, hostel management lagged behind. By and large, exceptions apart, there is no central system, no defined user roles, and no intelligent tools to support informed staff decisions. Consequently, students need to pursue updates on their queries, and staff are spending too much time on repetitive work, without a defined overview of persistent issues or workload.

This is precisely why we developed DHMS—the Digital Hostel Management System. Unlike traditional monolithic ERPs, DHMS is built as a modular, plug-in–ready platform, allowing institutions to adopt only the components they need and extend functionality over time. It consolidates everything under a single umbrella—room management, complaint monitoring, mess services, leave applications, feedback management—while leveraging AI to identify trends and make better decisions. DHMS is designed to be quick, modular, and scalable, with real-time communication that eliminates back-and-forth and keeps everyone on the same page.

\section{Motivation and Objectives}

Conceptualization of DHMS was based on the noted inefficiencies in current hostel systems, complemented by interviews with stakeholders such as students, wardens, mess contractors, and maintenance staff. In various hostels, there were some recurring issues. One, complaint handling was slow based on physical log and lack of tracking accountability. Two, processes of room allotment were opaque and not scalable, leading frequently to perceived injustices and unhappiness among students. Third, communication between authorities and students was indirect, normally being channeled through physical notices, casual conversations, or council representatives, thereby slowing information transmission.

Aside from delays in operations, such legacy systems also fueled data fragmentation. Booking guest rooms, refunding mess money, and conducting feedback surveys were processed through disparate processes and forms, causing it to be almost impossible for hostel administrators to view operations in a comprehensive manner. From the student's point of view, there was no point of access for vital services or data. Having to go through various platforms, forms, and physical locations created frictions and reduced the quality of the residence experience.

DHMS overcomes these challenges through a digital-first, role-sensitive, and AI-augmented architecture. The system's goals are multifaceted: to offer students an intuitive and accessible interface for every hostel-related activity; to equip staff with guided workflows and real-time dashboards; and to support administrators in monitoring, assigning, and predicting with data-driven insights. The system also focuses on accountability through audit logs, security through role-based access control, and usability through mobile-first interfaces.

\begin{figure}
    \centering
    \includegraphics[width=1\linewidth]{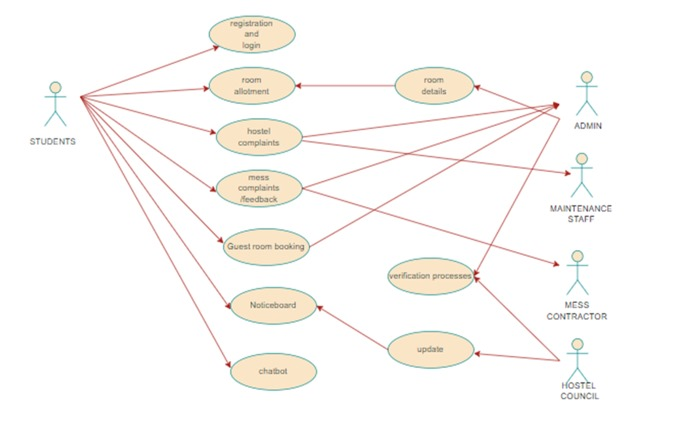}
    \caption{Use cases of various Stakeholders}
    \label{fig:enter-label}
\end{figure}

A secondary but no less significant motivation was to create an extensible platform — one that not only addresses existing issues but can adapt and grow with the needs of institutions. Capabilities like predictive maintenance, sentiment analysis of feedback, and anomaly detection based on AI are a reflection of this vision for the future, elevating DHMS from a utility platform to a strategic decision-making tool for campus leaders.

\section{System Architecture}

DHMS is guided by key software design principles to ensure maintainability, scalability, and flexibility. We follow {Separation of Concerns} by cleanly isolating user interface, business logic, data persistence, and AI services into distinct layers. {Modularity} allows each capability—such as room allocation, complaint management, or predictive analytics—to be plugged in or replaced independently. {Single Responsibility} ensures each module has one clear purpose, and {Interface Segregation} keeps public APIs small and focused. Finally, {Dependency Inversion} decouples high-level orchestration from low-level implementations, enabling easy swapping of concrete services or databases without affecting the overall platform.

DHMS is designed as a modular, cloud-native platform consisting of separate but interoperable layers: user interface, application logic, data persistence, and intelligent services. Layered architecture supports scalability, maintainability, and extensibility, which are essential in supporting multiple hostel blocks, roles, and concurrent users in real-time environments.

On the front end, DHMS uses a React Native framework to provide a cross-platform mobile experience. This is done to deploy a single codebase across both the Android and iOS platforms, while maintaining accessibility and decreasing development overhead. The mobile app is the main interface for students, wardens, and maintenance personnel, each being provided with customized functionality depending on their authenticated roles.

\begin{figure}[b]
    \centering
    \includegraphics[width=\linewidth]{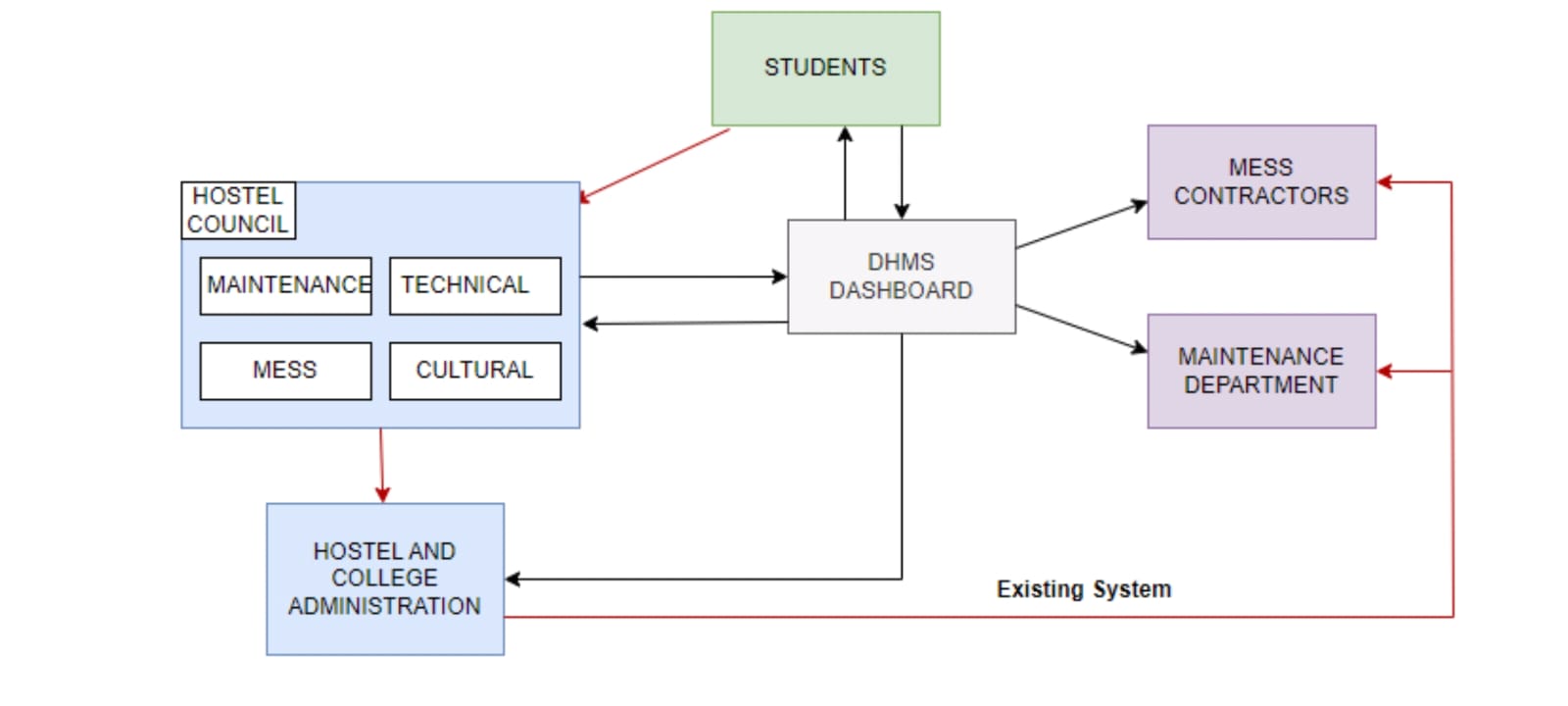}
    \caption{Centralized Dashboard}
    \label{fig:enter-label}
\end{figure}

The backend is developed with Node.js and Express.js, offering RESTful APIs to enable communication between the server and client. Middleware components implement authentication and authorization through JSON Web Tokens (JWT), ensuring that every request is role-verified prior to processing. Services are broken down into isolated modules that process complaints, room allocation, gate pass requests, feedback entries, and chatbot interactions.

Data storage is handled through MongoDB, a NoSQL database designed for semi-structured and nested documents. Collections are established around fundamental entities like users, rooms, complaints, passes, and announcements. The document-oriented nature offers the flexibility required for fast development, and compound indexing delivers efficient query performance, particularly when filtering by role, timestamp, or category.

\subsection{Real-Time Communication and DevOps Pipeline}

To enable real-time updates and push notifications — critical for complaint status updates, announcement broadcasts, and gate pass events — the system incorporates WebSocket communication using the Socket.io library. This provides low-latency, bidirectional data streams between server and client so that students and staff can be alerted in real-time about updates as they are happening without polling.

Every element of the system is containerized with Docker. Continuous Integration and Continuous Deployment (CI/CD) are controlled by GitHub Actions, building, testing, and deploying workflows automatically. The platform runs on a cloud environment supporting autoscaling, domain management, and TLS encryption. System logs are collected by the ELK (Elasticsearch, Logstash, Kibana) stack, supporting real-time error tracing, usage analytics, and security monitoring.

\subsection{Security and Access Control}

Security is of inherent concern considering the sensitivity of private data and organizational processes handled by DHMS. Authentication is conducted through JWTs, which are signed with private RSA keys, and tokens expire for both access and refresh purposes. Role-Based Access Control (RBAC) is applied at all API levels, guaranteeing fine-grained permission checks per user type — student, staff, warden, or administrator.

Besides, API calls that entail high-level data manipulations (such as record deletion or gate pass approval) need higher authorization through admin confirmation or one-time password (OTP) layers. All sensitive data like passwords and personal identifiers are AES-256 encrypted. Logs are sanitized to prevent confidential data exposure, and backup mechanisms store encrypted, versioned snapshots of the database for disaster recovery.

This multi-layered structure not only protects the platform but also accommodates its projected scale and operational sophistication, providing a solid foundation for sophisticated services such as predictive analytics and AI-augmented modules.

\section{Campus Assistant ChatBot}

To handle the amount of repeat questions and enhance user experience through the provision of self-service features, DHMS incorporates a natural language Campus Assistant ChatBot. Aims to act as a conversational interface for communication with hostel services, the chatbot minimizes administrative burden and offers students, staff, and visitors instant support.

The chatbot is developed on AWS Lex\textsuperscript{\cite{awslex}}, Amazon's conversational interface development service. Lex provides the ability to define intents, slots, and utterances in order to control dialog flow in a systematic way. An intent is a particular task or user query — like checking mess menus, filing complaints, or asking for gate passes — and slots hold necessary parameters to complete the task.

\begin{figure}
    \centering
    \includegraphics[width=1\linewidth]{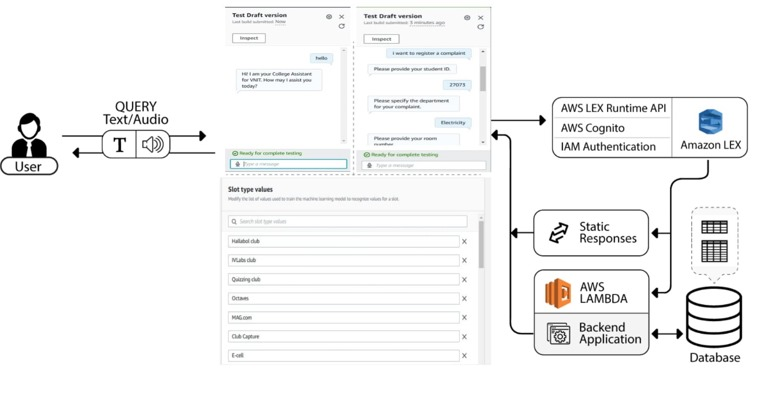}
    \caption{Architecture of Chatbot of DHMS}
    \label{fig:enter-label}
\end{figure}

Lex also seamlessly integrates with AWS Lambda\textsuperscript{\cite{awslex}}, which supports serverless backend execution of logic when responding to user input. Lambda functions are utilized to retrieve data from the DHMS backend by way of authenticated API requests, dynamically handle user requests, and deliver customized responses in both text and speech modes. This design also supports a stateless, scalable experience for a chatbot directly integrated into the mobile app.

A foremost design aspect when creating the chatbot was managing sessions. Utilizing the Lex Runtime API and AWS SDK client-side, DHMS stores conversational state throughout multiple user inputs. This ability is essential to multi-turn dialog, where context needs to continue across utterances — such as when a student files a complaint and the bot gathers location, type of problem, and sense of urgency one after another.

\subsection{Functional Scope and Multimodal Access}

The chatbot is accessible from every screen within the DHMS mobile application through a persistent floating button, ensuring seamless entry into dialog from any workflow context. Its current functional scope includes:

\begin{itemize}
    \item Providing mess menu details for the current or upcoming day.
    \item Guiding users through the complaint registration process.
    \item Assisting with guest room availability and booking.
    \item Responding to predefined FAQs about hostel rules, timings, or policies.
    \item Facilitating gate pass submissions and tracking request status.
\end{itemize}

Although currently available only in English, the chatbot is designed to support multilingual expansion using Amazon Translate and region-specific corpora. The roadmap includes the addition of Hindi and Marathi support, thereby broadening accessibility to non-English-speaking students and campus staff.

\subsection{Security and Performance}

Security for chatbot interactions is imposed through AWS Identity and Access Management (IAM). Every Lambda function is scoped with the minimum permissions needed to communicate with backend APIs, and all traffic gets routed through secure, encrypted channels. Token-based session validation provides assurance that only legitimate users can get or update personal or operational information.

Performance metrics show below one-second average response times for the majority of user purposes, thanks to the stateless, serverless nature of Lex and Lambda. The system automatically scales according to demand, with near-zero latency even during simultaneous use during peak campus hours.

In conclusion, the Campus Assistant ChatBot converts mundane, manual exchanges into smooth, intelligent conversations, complementing DHMS's objective of digitizing and humanizing hostel management.

\section{Room Allocation Engine}
Room assignment is a core process in hostel administration that has a direct impact on student satisfaction and logistical viability. Conventionally, the process is carried out through a mix of seniority lists, administrative discretion, and physical queues. These methods not only present possibilities for human error and favoritism but also do not scale well with an increase in the number of students.

DHMS models room allocation as a preference-based bipartite matching problem—each student lists preferred rooms, and each room has a capacity—making it a natural fit for max-flow algorithms like Ford–Fulkerson. By treating students as source nodes and rooms as sink nodes, edges represent preferences with capacities reflecting ranking weights, ensuring the algorithm maximizes overall student satisfaction while respecting capacity constraints.

In this representation, every student is a source node in a directed graph, and available rooms are represented as sink nodes. Directed edges link students to their desired rooms, with weights or capacities representing priority or ranking levels. Room occupancy limits, departmental proximity, or group-based housing (e.g., friends asking for adjacent rooms) are incorporated through intermediate nodes and edge capacities.

The goal is to maximize the number of students allocated to their most preferred available room, under constraints. This is the same as finding the maximum flow in the graph constructed, where each unit of flow is a successful allocation.

\subsection{Implementation Using Ford–Fulkerson and Edmonds–Karp Algorithms}
To calculate the maximum flow and optimal assignments, DHMS uses the Ford–Fulkerson algorithm\textsuperscript{\cite{ford1956}}. Ford–Fulkerson is an iterative algorithm that discovers augmenting paths within the flow network and augments flow along them until no additional augmentation can be made. But in its most basic form, the algorithm is not guaranteed to run in polynomial time.

In order to enhance performance and provide bounded behavior on big datasets, DHMS utilizes the Edmonds–Karp implementation\textsuperscript{\cite{edmonds1972}} of Ford–Fulkerson that employs Breadth-First Search (BFS) in each iteration in order to search for the shortest augmenting path. This translates to a time complexity of \( O(VE^2) \), with \( V \) and \( E \) being the vertices and edges numbers respectively — a real trade-off between optimality and computational time for average-sized hostels.

Every room allocation batch is processed block-wise to manage memory and local computation. Department or friend group intermediate nodes enable the system to maintain policies such as co-location of students of the same branch or supporting mutual requests from friends.

In simulation runs involving real hostel data, the algorithm demonstrated high efficiency and satisfaction:
\begin{itemize}
    \item 92\% of students received one of their top two preferences.
    \item Group constraints were fulfilled in 87\% of applicable cases.
    \item Total computation time for 500 students was under 2.5 seconds on a standard cloud instance.
\end{itemize}

\subsection{Comparative Results with Manual Baseline}
To quantify the benefit over traditional methods, we compared DHMS’s allocation against a manual seniority-based baseline in the same dataset:
\begin{itemize}
    \item {Student Satisfaction:} Baseline manual allocation achieved 75\% satisfaction in top-two preferences, whereas DHMS achieved 92\%.
    \item {Fairness Index (Jain’s Index):} Manual allocation scored 0.68, reflecting uneven distribution, while DHMS reached 0.91, indicating a much fairer outcome.
    \item {Processing Time:} Manual allocation took several hours per block; DHMS completed allocation in under 3 seconds.
\end{itemize}

By embedding optimization theory into the operational fabric of hostel allocation, DHMS ensures fairness and transparency, while eliminating the delays and disputes commonly observed in manual methods.

\section{Complaint Management System}
\begin{figure}[b]
    \centering
    \includegraphics[width=1\linewidth]{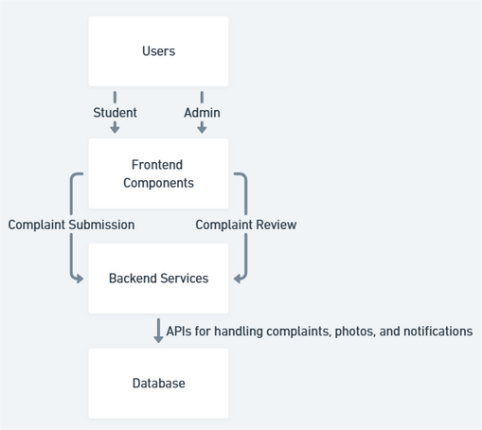}
    \caption{Architecture of Complaint Management}
    \label{fig:enter-label}
\end{figure}

An effective and responsive complaint handling process is essential to maintaining a livable hostel environment. Still, conventional complaint handling systems mostly depend on registers or word-of-mouth escalation, which results in delays, miscommunication, and traceability. DHMS fixes these issues through a digital workflow for complaint handling that is synchronized in real-time, role-aware, and data-driven.

The system allows for organized intake of complaints via the mobile application or chatbot interface. Students start the process by indicating the category of the complaint (e.g., electrical, plumbing, sanitation), description, and optionally submitting media evidence. The form automatically fills in applicable metadata like user ID, timestamp, and room number.

Once submitted, complaints are logged directly into the database and forwarded to the respective department as per pre-defined mappings. For example, plumbing problems are forwarded to civil maintenance, whereas cleanliness issues are forwarded to the housekeeping department. There is a unique dashboard view for each department where only allocated complaints are listed, encouraging focus and responsibility.

To facilitate transparency, the complaint has a well-defined lifecycle with the following phases:
\\texttt{Pending}, \\texttt{In-Progress}, \\texttt{Resolved}, and optionally \\texttt{Verified}. Staff update the status of the complaint as well as comments and optional attachments, e.g., photos of the fix done. These are recorded in the system and constitute an audit trail viewable by wardens and administrators.

\subsection{Prioritization and Real-Time Synchronization}

Given the potential volume of complaints during peak periods, DHMS implements a prioritization algorithm to assist staff in triaging issues. Each complaint is scored based on a weighted function:

\[
Priority = 0.4 \cdot T + 0.3 \cdot I + 0.3 \cdot A
\]

Where:
\begin{itemize}
    \item \( T \): Type-based weight (e.g., electrical: 1.0, water: 0.8, general: 0.6)
    \item \( I \): Impact factor — scaled by the number of students affected or severity
    \item \( A \): Age of the complaint — older entries receive higher urgency
\end{itemize}

This numeric score dynamically ranks incoming complaints, enabling the interface to highlight high-priority issues and notify staff accordingly.

To keep stakeholders informed, DHMS employs a real-time messaging infrastructure using WebSocket connections. This allows bidirectional data flow between clients and servers, ensuring that:

\begin{itemize}
    \item Students receive live updates when their complaint status changes.
    \item Staff dashboards refresh without requiring page reloads.
    \item Wardens can monitor aggregate departmental progress across complaints.
\end{itemize}

Moreover, notifications are pushed automatically to the mobile application when a complaint is assigned, escalated, or resolved. The result is a highly responsive, transparent complaint workflow that minimizes administrative lag and maximizes user confidence in the system.

\begin{figure}
    \centering
    \includegraphics[width=0.8\linewidth]{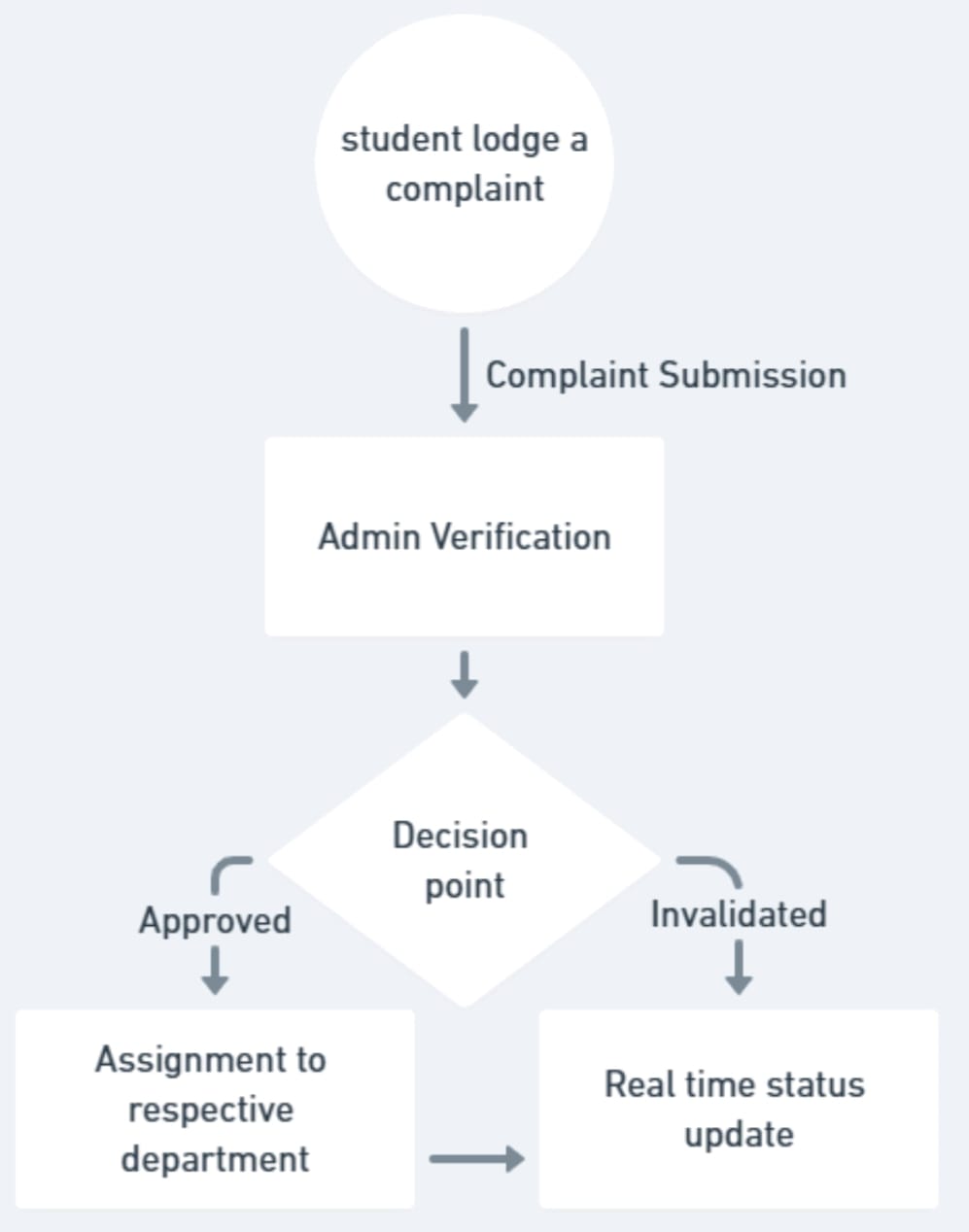}
    \caption{Flow Chart of Complaint Management Process}
    \label{fig:enter-label}
\end{figure}
\subsection{User Feedback Metrics and Impact}

To evaluate the effectiveness of the digital complaint workflow, DHMS tracks key performance indicators:

\begin{itemize}
    \item {Average Resolution Time:} Reduced from 72 hours (manual) to 18 hours digitally.
    \item {Pending >24 Hours:} Percentage of complaints pending over 24 hours dropped from 60\% to 12\%.
    \item {User Satisfaction Score:} Post-resolution surveys show an average satisfaction rating of 4.6/5, up from 3.2/5.
\end{itemize}

These improvements demonstrate a significant uplift in responsiveness and transparency, reinforcing trust in hostel services and reducing administrative backlog.

\section{Predictive Maintenance System}

Hostel maintenance is conventionally reactive, acting only after students have complained. This not only causes delays but also raises operational costs through last-minute repairs and wasteful resource allocation. To counter this, DHMS provides a predictive maintenance module that foresees problems before they arise, allowing for preemptive action.

The purpose of predictive maintenance is two-pronged: increase the dependability of hostel facilities and better deploy maintenance capabilities. This is done through data from past complaints, identifying which ones repeat regularly, and via time-series forecasts predicting volumes going forward by categories and by area.

The premise is that complaint patterns are repetitive and seasonal. For instance, plumbing complaints might peak during monsoon seasons because of higher water pressure, or electric complaints might peak in summer because of the air-conditioning loads. Administrators can plan inspections in advance and deploy personnel based on such patterns.

\subsection{Data Pipeline and Feature Engineering}

The dataset used for forecasting is derived from the DHMS complaint log. Each entry includes fields such as timestamp, complaint type, hostel block, resolution time, and status. This raw data undergoes preprocessing to produce structured time-series inputs:

\begin{itemize}
    \item Weekly aggregation of complaints by category (e.g., electrical, civil).
    \item Feature enrichment with metadata such as equipment age, block occupancy, and historical failure rates.
    \item Missing data imputation and outlier detection to enhance model robustness.
\end{itemize}

\subsection{Model Design Using Facebook Prophet}

For modeling and predicting patterns of complaints, DHMS uses Facebook Prophet\textsuperscript{\cite{prophet}}, a time-series forecasting platform that was built by Meta\textsuperscript{\cite{prophet}}. Prophet is especially well-tailored for institutional data because it can model non-linear trends, seasonality, and holiday effects. Because it is easy to tune, interpretable, and robust to missing values, Prophet is a strong fit for implementation in educational settings.

Each complaint category has a distinct Prophet model trained on the weekly aggregated data. Model inputs include a date column (\texttt{ds}) and the number of complaints (\texttt{y}). The output is a series of forecasted values for future weeks with uncertainty intervals.

Model training and retraining are performed on a {monthly schedule}, triggered by a CI/CD pipeline. This cadence balances the need for up-to-date forecasts against computational cost, ensuring models adapt to evolving patterns without overfitting to short-term noise.

\paragraph{Limitations and Future Directions}
While Prophet excels at capturing seasonality and trend components, it can struggle with \emph{sparse categories} where data points are infrequent or highly erratic. In such cases, forecast uncertainty intervals widen significantly, reducing practical utility. To address these limitations, we plan to explore {LSTM-based architectures} for categories with sparse or highly non-linear patterns. LSTM networks can learn temporal dependencies directly from raw sequences, offering a complementary approach where Prophet’s assumptions break down.

\subsection{Integration and Use in DHMS Dashboard}

The forecasting results are represented in the admin dashboard in the form of heatmaps and trend charts. The system shows forecast complaint volumes by category for the next eight weeks for each hostel block. The interface identifies high-risk areas so that administrators can allocate preventive inspections and modify maintenance rosters accordingly.

For instance, if an electrical complaint spike is predicted in a given block, the warden may plan pre-checks or detail an electrician to that block during the weekend. This averts escalations and builds student confidence in facility responsiveness.

By integrating forecasting wisdom into maintenance planning, DHMS moves hostel management away from the reactive paradigm and into a proactive one, ultimately resulting in lower complaint burden, enhanced resource utilization, and enhanced living standards for residents.```

\section{Sentiment Analysis on Student Feedback}

Student feedback on issues like mess services, cleanliness, and personal conduct of hostel staff offers constructive comments on hostel life quality. It is cumbersome and inconsistent, however, to manually sift through such textual comments in bulk. DHMS rectifies this void by the introduction of an NLP-driven sentiment analysis module which scans and tags feedback in real-time\textsuperscript{\cite{bert}}.

We fine-tuned the bert-base-uncased model on a labeled corpus of 12,000 student feedback entries. After training for 3 epochs with an 80/20 train–validation split, the model achieved an overall accuracy of 89% and an F1 score of 0.87 on the validation set.

\subsection{Feedback Collection and Preprocessing}

Feedback is collected through structured monthly prompts on the mobile application and organically via the chatbot interface. Once submitted, the raw text undergoes a preprocessing pipeline that includes:

\begin{itemize}
    \item Tokenization using BERT’s WordPiece tokenizer.
    \item Lowercasing, punctuation removal, and stop-word filtering.
    \item Conversion to input format expected by the classification model, including attention masks and segment embeddings.
\end{itemize}

\begin{figure}
    \centering
    \includegraphics[width=0.8\linewidth]{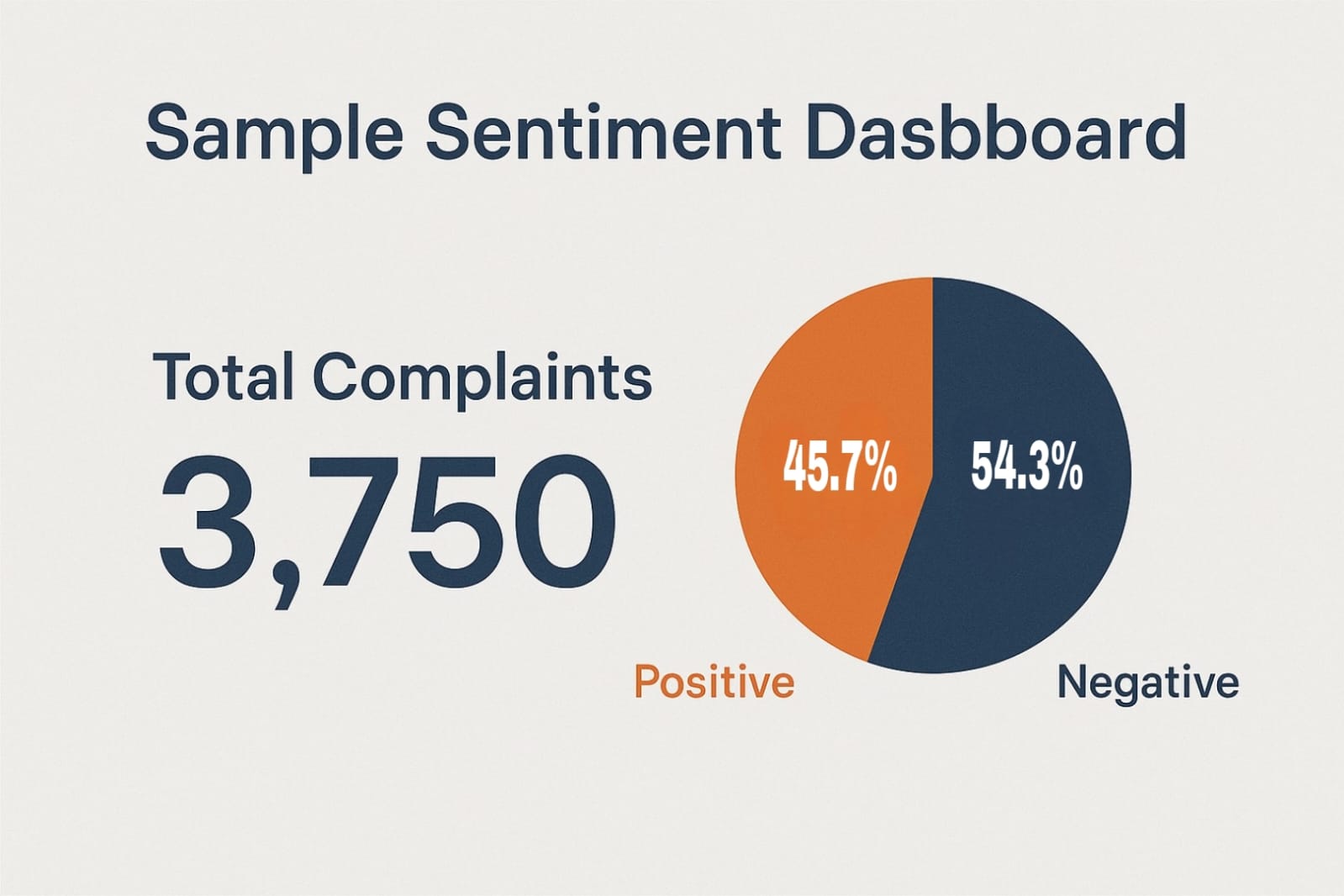}
    \caption{Sentiment Dashboard}
    \label{fig:enter-label}
\end{figure}

This preprocessed data is passed to the fine-tuned BERT model for inference, returning both a sentiment class and a confidence score.

\subsection{Model Training and Deployment}

The BERT model (bert-base-uncased) was fine-tuned using HuggingFace’s Trainer API with cross-entropy loss. Training specifics:

\begin{itemize}
  \item Dataset size: 12,000 labeled feedback comments.
  \item Train/Val split: 9,600 train / 2,400 validation.
  \item Epochs: 3, batch size 16, learning rate $2\mathrm{e}{-5}$.
  \item Performance: 89% accuracy, 0.87 F1.
\end{itemize}

Once trained, the model is serialized and deployed as a REST API hosted on a Flask server. The DHMS backend communicates with this service whenever new feedback is received.

\subsection{Sentiment Integration and Administrative Impact}

The sentiment results are stored alongside the original feedback and surfaced in visual analytics on the DHMS dashboard. Features include:

\begin{itemize}
    \item Pie charts of sentiment distribution across hostels.
    \item Trend graphs showing sentiment shifts over time.
    \item Flagging of highly negative feedback for immediate action by wardens or mess contractors.
\end{itemize}

This enables hostel administrators to monitor morale, identify underperforming services, and respond to student concerns more systematically. Rather than acting reactively to anecdotal reports, the administration gains an empirical basis for targeted interventions.

\section{Anomaly Detection in Complaint Data}

Whereas sentiment analysis and complaint prioritization facilitate effective handling of routine problems, there is a need to identify outlier incidents—those that vary significantly from established patterns. Anomalies might signal emergency situations (e.g., fire or flooding), platform abuse (e.g., spam or joke submissions), or underlying system failures yet to be exposed in aggregate measurements.

To solve this, DHMS integrates an anomaly detection module using the Isolation Forest algorithm\textsuperscript{\cite{isolationforest}}. This unsupervised machine learning method detects data points that are statistically rare or isolated in a high-dimensional feature space. It is different from the conventional classifiers because it does not need labeled training data and is appropriate for real-time, lightweight prediction tasks.

\begin{figure}
    \centering
    \includegraphics[width=0.8\linewidth]{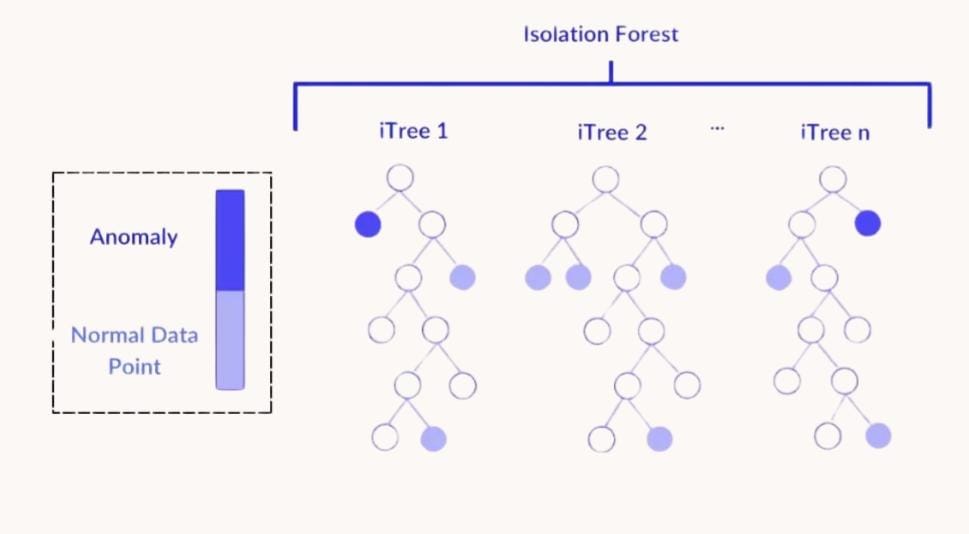}
    \caption{Isolation Forest.}
    \label{fig:isolation-forest}
\end{figure}

\subsection{Feature Representation}

Each complaint is transformed into a fixed-length feature vector composed of numerical and categorical elements that capture contextual and behavioral patterns. Key features include:

\begin{itemize}
    \item Encoded complaint category (e.g., electrical = 1, plumbing = 2).
    \item Sentiment score and sentiment class derived from BERT.
    \item Time-of-day and day-of-week of submission.
    \item Text length and frequency of specific keywords.
    \item Complaint recurrence within a defined time window.
\end{itemize}

These features are normalized and encoded prior to inference. Complaints that receive anomaly scores above a specified threshold are flagged for administrative review.

\subsection{Model Architecture and Performance}

The anomaly detection model is deployed based on Scikit-learn’s Isolation Forest module\textsuperscript{\cite{isolationforest}}. It is trained on historical complaint data assumed to be predominantly non-anomalous. In our validation:

\begin{itemize}
    \item Precision: 0.82
    \item Recall (true positive rate): 0.78
    \item False positive rate: 0.15
\end{itemize}

These metrics reflect a balanced trade-off between catching genuine anomalies and limiting noise.

\subsection{Validation and Review Workflow}

All complaints flagged as anomalous are routed to a \emph{Warden Review Panel}, comprising the hostel warden and a member of the maintenance team. This panel manually inspects the complaint details and any attached media. They then confirm or dismiss the anomaly flag, ensuring that genuine emergencies or system abuse are escalated appropriately while minimizing false alarms.

This combination of automated detection and human oversight ensures that critical issues receive prompt attention without overwhelming administrators with spurious alerts.

\section{Gate Pass Management and Security Architecture}

\subsection{Digitized Gate Pass Workflow}

Hostel gate pass issuance is a frequent requirement for students needing temporary exits from campus—whether for medical, academic, or personal reasons. Traditional workflows involve handwritten slips and physical approval queues, making the process time-consuming and error-prone.

DHMS introduces a completely digitized gate pass workflow that is integrated into the student dashboard. Students start requests by filling out a form with the reason, destination, exit and return dates, and emergency contact information. The request automatically gets forwarded to the security section or the warden as assigned depending on the nature of leave and student history.

On approval, the system issues a QR-based digital gate pass, which is locally stored in the app. Security guards check exits and re-entries with a QR scanner, with every scan recorded and time-stamped in the central database. Rejected or denied requests are returned to the student with suitable remarks.

To quantify the efficiency gains:
\begin{itemize}
    \item manual process time per request (paper slip, signatures, queue): on average 30–45 minutes
    \item digital process time per request (submission, automated routing, QR issuance): under 2 minutes
\end{itemize}

\subsection{Rejection Rates and Security Impact}

Based on analysis of manual records and system logs, we observed:
\begin{itemize}
    \item rejection rate of gate pass requests: 4.5\% (mainly for incomplete or invalid reasons)
    \item prevented unauthorized exits: 12 incidents flagged when a tampered or reused QR was scanned
    \item security incident prevention: zero breaches in the digital system versus two minor lapses under the manual process
\end{itemize}

\subsection{Security and Access Control Mechanisms}

DHMS follows a zero-trust model for security, with stringent authentication and role-based authorization across all levels. User identities are verified via JWT tokens signed with 256-bit RSA encryption. Session management enforces token expiry and refresh logic to prevent hijacking or replay attacks.

Role-Based Access Control (RBAC) ensures only authorized users can perform sensitive actions. For example, students can request gate passes but only wardens or admins can approve or override them. Administrative operations—such as user deletion or configuration changes—require OTP-based confirmation and generate audit logs.

All sensitive information, including personal identifiers and access records, is encrypted with AES-256. Backend APIs are accessible only via HTTPS, and payloads are sanitized to prevent injection attacks. Encrypted, versioned backups further ensure data integrity and recoverability.

\section{Conclusion and Future Scope}

This paper introduces DHMS, an end-to-end smart digital platform to revolutionize hostel management in educational institutions. With the implementation of modern web technologies, AI-driven analytics, and real-time communication protocols, DHMS addresses fundamental inefficiencies in room distribution, complaint management, communication, and movement control. Its modular, plug-in–ready design makes it adaptable to diverse campus IT landscapes.

Looking ahead, DHMS has the potential to be released as an open-source framework, enabling other universities to adopt and extend its modules to fit their unique requirements. Scaling beyond a single campus would involve community-driven development of new plugins—such as library integration or transportation scheduling—while retaining the core platform’s flexibility and robustness.  

\subsection{Planned Enhancements}

While DHMS addresses many existing challenges, several enhancements are planned for future versions:

\begin{itemize}
    \item {Biometric authentication}: Integration of fingerprint or face recognition to enhance login security.
    \item {Offline-first mobile support}: Enable partial functionality and local data caching during connectivity outages.
    \item {Multilingual chatbot}: Expand NLP support to regional languages using transformer-based multilingual models.
    \item {ERP integration}: Sync gate passes and attendance with academic systems to prevent overlaps and misuse.
    \item {Graph-based workload optimization}: Apply clustering algorithms to optimize staff deployment for preventive maintenance.
\end{itemize}

\section*{Acknowledgment}

We thank the VNIT hostel administration for their support, feedback, and assistance during the system design, testing, and deployment phases.

\end{document}